# Vehicular Edge Cloud Computing: Depressurize the Intelligent Vehicles Onboard Computational Power

Xin Li
School of Information Science & Engineering
Central South University
Changsha, China
lixin9206@gmail.com

Yifan Dang
University of Oregon
Eugene, OR, USA
yifand@uoregon.edu

Tefang Chen
School of Information Science & Engineering
Central South University
Changsha, China
ctfcyt@163.com

*Abstract*—Recently, with the rapid development of autonomous vehicles and connected vehicles, the demands of vehicular computing keep continuously growing. We notice a constant and limited onboard computational ability can hardly keep up with the rising requirements of the vehicular system and software application during their long-term lifetime, and also at the same time, the vehicles onboard computation causes an increasingly higher vehicular energy consumption. Therefore, we suppose to build a vehicular edge cloud computing (VECC) framework to resolve such a vehicular computing dilemma. In this framework, potential vehicular computing tasks can be executed remotely in an edge cloud within their time latency constraints. Simultaneously, an effective wireless network resources allocation scheme is one of the essential and fundamental factors for the QoS (quality of Service) on the VECC. In this paper, we adopted a stochastic fair allocation (SFA) algorithm to randomly allocate minimum required resource blocks to admitted vehicular users. The numerical results show a great effectiveness of energy efficiency in VECC.

## I. INTRODUCTION

In recent years, a new trend in cloud computing has been increasingly moving to the network edge, named as mobile edge computing (MEC) by the European Telecommunications Standards Institute (ESTI) [1]. Comparing to the conventional cloud computing, mobile edge cloud computing can be widely distributed at the edge of the network, such as the E-UTRAN Node Base station (eNodeB). The data between the edge cloud server and edge devices can be transmitted at a much lower latency than in cloud computing [2] [3].

Meanwhile, a great interest in autonomous vehicles and connected vehicles is raising in both academia and industry. These intelligent vehicles are assembled with various sorts of modules, such as sensors (e.g. radar, camera, Lidar, IMU, DGPS), communication systems (e.g. DSRC, 4G, WIFI) and onboard computing units (e.g. CPU and GPU). An ocean of computing operations are executed locally (i.e. on vehicles) nowadays. It is not hard to predict that the demands for embedding interactive applications and computing resource harvest services on intelligent vehicles will keep increasingly growing for decades. However, a constant vehicular onboard computing ability can hardly keep up with the growing speed of computing harvest demands, which will indeed become a big challenge for the old model intelligent vehicles. Furthermore, autonomous vehicles are currently facing a potential problem which will cost higher energy consumption [4] [5]. Last but not the least, the abundant vehicular modules are also tending to be engaged in management, maintenance and record over their lifetime.

In this case, we propose a Vehicular Edge Cloud Computing (VECC) framework in this paper, which fundamental idea is offloading vehicular computing operations to the edge cloud through the wireless network. In this approach, partial or entire of the computing operations on the vehicles can be migrated (or offloaded) to scalable and distributed managed resource pools. Each vehicle can flexibly request for its temporarily individual CPU, GPU and storage resources during its operations.

In the literature, there are some recent efforts dedicated to related vehicular cloud computing and the network resource allocation. In [6], the authors investigated how to provide the time constraint computing services in vehicular cloud computing system and proposed a balanced-task-assignment policy to minimize the deadline violation probability of offloading vehicular computation tasks. A vehicular fog cloud computing architecture was proposed in [7], the authors emphasized on the cooperation among multiple edge servers, which can provide uninterrupted vehicular services during vehicle movement. The authors in [8] considered the security issues for the vehicular edge computing and proposed a reputation-based resource allocation scheme for different reputation values vehicular users. More impressively, the authors in [9] proposed a vehicular cloud computing prototype to place and schedule the application workloads and had a field experiment to evaluate the performance of vehicular cloud computing in Pittsburgh. The results show effectiveness in reducing the operation time by vehicular edge cloud computing. In [10] [11], the authors studied the wireless resource allocation strategy in a multi-user and multi-access scheme's scenario. Their research focused on optimizing the end user's energy efficiency in offloading computation. In a different perspective, the authors in [12] studied the wireless network resource allocation strategy in a multi-user and multi-access scheme's scenario. Their research focused on optimizing the end user's energy efficiency in offloading computation. They aimed at minimizing the total consumption of both energy and time in edge computing offloading.

In order to achieve an efficient vehicular edge cloud computing offloading service for multiple vehicular users, we need to figure out two principal issues: 1) how should a vehicular user make a decision between local computing and offloading computing? 2) how should an eNodeB controller allocate a reasonable amount of wireless bandwidth to each offloading user?

In this paper, we proposed a both time and energy aware vehicular computing offloading framework with time latency constraints, which can be adapted to the Internet of Vehicles (IoV) research works. Furthermore, a low computational complexity algorithm for stochastically and fairly allocating the wireless bandwidth resources among all of the vehicular users are proposed. It shows us the benefits of VECC in specific.

The rest of this paper is organized as follows. Section II describes the resource allocation process model within our research framework. Section III proposes the optimizing issues and allocation algorithm. Evaluation of VECC framework and resource allocation algorithm are analyzed in section IV. Conclusion is in the final section V.

## II. VEHICULAR EDGE CLOUD COMPUTING FRAMEWORK AND SYSTEM MODEL

In this section, we specifically introduce our vehicular edge cloud computing framework and the mathematical model of the resource allocation process.

As illustrated in Figure 1, a single eNodeB, equipped with an edge cloud server, is serving for the intelligent vehicles in its wireless radiation area. The intelligent vehicles need to be equipped with the wireless communication system, so they can be served by the edge cloud services. Considering these vehicles involve different computing tasks and are equipped with diverse hardware and system (or software). Their onboard computing ability and power consumption can be separated into multiple levels. Therefore, the arguments of different vehicular computing tasks can be further separated into distinct time consumption, energy consumption, and time latency constraint. If the offloading computing time consumptions were met with the demands of latency constraints, the actual vehicular computing tasks could be implemented in our VACC framework, theoretically. Such as the huge number of deep learning-based inference applications for nowadays autonomous driving system. Moreover, one or more times handovers might happen during vehicles movements, which will involve other complex issues (e.g. data transfer and computing migration). Therefore, we will involve the handover process in our next step research.

We can model the vehicular users set as $N = \{1, 2, ..., N\}$. The vehicular computing task of vehicular user $i$ can be noted as $\Gamma_i \triangleq \{\delta_i, \beta_i, \gamma_i\}$, where $\delta_i, \beta_i$ and $\gamma_i$ denote the size of computing input data (e.g. the program codes or sensors data sets), the total required computation units (e.g. the CPU Cycles in CPU computation or the Flops in GPU computation) and the time latency limit required to accomplish the computation task, respectively.

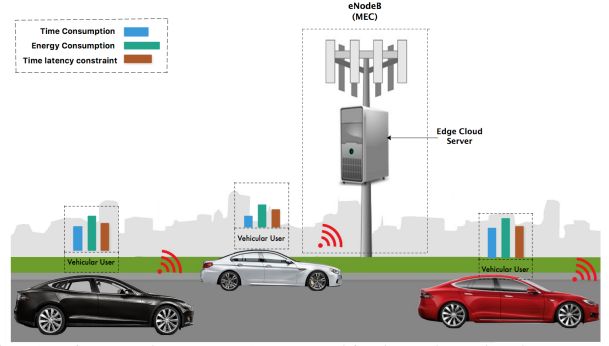

Fig. 1: Time and Energy Aware Vehicular edge cloud computing framework

The task set $\Gamma$ can be executed either locally or remotely. When task $i$ is executed locally (i.e. vehicular onboard computer), the execution overhead of task $i$ consists of local energy consumption $e_i^l$ and local execution time $t_i^l$. We can compute the energy consumption and execution time of user $i$ by the following equations.

$$e_i^l = \varphi \beta_i \tag{1}$$

$$t_i^l = \frac{\beta_i}{f_i^l} \tag{2}$$

The overall overhead for local computing of user $i$ is:

$$O_i^l = \lambda_i^t t_i^l + \lambda_i^e e_i^l \tag{3}$$

$\lambda_i^t$ and $\lambda_i^e$ are the weights of execution time and energy consumption, separately. We have $\lambda_i^t + \lambda_i^e = 1$ for each users computation task. For example, when a user set $\lambda_i^t = 1$ and $\lambda_i^e = 0$ in the decision making, it means the user only cares about time latency, vice versa. Correspondingly, if a user set $\lambda_i^t = 0.5$ and $\lambda_i^e = 0.5$, that means this user cares about both of the time latency and energy consumption, equally. According to the different priority configuration of energy and time consumption of diverse users and applications, we can set any decimal for $\lambda_i^t$ and $\lambda_i^e$ in the range of 0 to 1, as long as their sum equals to 1.

As for the edge cloud computing. The wireless network prototype in our research utilizes currently widely used 4G network. The eNodeB is a base station which manages abundant uplink/downlink data communications. For considering the capacity of the downlink is much higher than the uplink in the cellular network [13], and the computation results are much smaller than the computation input data (e.g. program code, sensing data sets) in most cases. Therefore, we only consider the uplink process in our model. Such a method is widely adopted in related research works [9] [10] [14].

We build a decision profile $A = \{a_1, a_2, ..., a_n\}$ for all the vehicular users, where $a_i$ denotes the amount of allocated resource block for vehicular user $i$. Specifically, if $a_i = 0$, it means user $i$ chooses compute the task locally on the vehicular by on-board computational devices (e.g.

CPU or GPU); otherwise, if $a_i > 0$, it denotes user $i$ offload the computation to edge cloud via $a_i$ resource blocks bandwidth. We can get the bandwidth and the uplink data rate of vehicular user $i$ based on Shannon-Hartley capacity by equation (4) and (5), separately:

$$B_i = 180 \ a_i \ (kHz) \tag{4}$$

$$r(a_i) = \sigma B_i \ log_2\left(1 + \frac{P_i|H_{i,s}|^2}{N_0}\right) \tag{5}$$

In formula (4), $B_i$ denotes the initial individual valid transmission bandwidth. Since each resource block consists of 12 subcarriers, and each subcarrier is allocated with 15 kHz bandwidth [15], we can calculate that each resource block occupies 180 kHz bandwidth. In equation (5), $\sigma$ denotes the number of carriers involved in the Carrier Aggregation technology. $P_i$ denotes user $i$'s transmission power on the allocated sub-channel, which is distributed by power allocation mechanism at the base station. $H_{i,s}$ is the channel gain between the vehicular user $i$ and the base station $s$, $N_0$ is the white Gaussian channel noise power. Because the uplink multi-access scheme of 4G cellular network uses the SC-FDMA technology, which belongs to the orthogonal multiple sub-channels communication. We herein do not take the channel interference among each user into our consideration.

The computing overhead of task $\Gamma_i$ for edge cloud computing also consists of two main parts, time and energy consumption. The $e_i^o$ denotes the energy consumption of data uploading transmission and $\tau_i$ denotes a tail energy which spent on continuously holding the channel after data offloading. The execution time consists of transmission time $t_i^{o,t}$ and execution time $t_i^{o,e}$ on the edge cloud. We can get the transmission time, execution time and energy consumption of user $i$ with equation (6)-(8), separately.

$$t_i^{o,t} = \frac{\delta_i}{r(a_i)} \tag{6}$$

$$t_i^{o,e} = \frac{\beta_i}{f_i^o} \tag{7}$$

$$e_i^o = \frac{\delta_i P_i}{r(a_i)} + \tau_i \tag{8}$$

The $f_i^o$ in formula (7) denotes the computation capability that the edge cloud allocated for user $i$. According to the 3 equations above, we can compute the overhead via computing offloading with equation (9).

$$O_i^o = \lambda_i^t(t_i^{o,t} + t_i^{o,e}) + \lambda_i^e e_i^o \tag{9}$$

Based on the system model above, we will analyze the performance differences between local computing and computing offloading in the next section and an efficient algorithm for the wireless resource allocation will also be proposed.

## III. VEHICULAR EDGE CLOUD COMPUTING OFFLOADING STRATEGY

In this section, we propose a system optimizing problem and a stochastic fair allocation algorithm for wireless resource allocation for the VECC.

In the local-offloading computing trade-off, choosing the smaller overhead approach is an optimal decision. The overhead can be influenced by a variety of elements (e.g. computing ability, time tolerant limit, wireless resource and energy consumption). Furthermore, in the practical circumstances, the capability of computation and storage in edge cloud is scalable. It can be upgraded and managed in a higher QoS than local computing. Therefore, the offloading computing approach will be recommended firstly when the computing requirements can be matched.

However, as we all know, the wireless resource pool isnt unlimited. When the amount of users grows, we need to figure out how to efficiently allocate a reasonable wireless resource among all of the requested users, simultaneously. This is a significant problem need to be resolved sooner or later. Moreover, the optimization of system performance is also a significant issue in VECC.

There are several thoughts to solve such issue (e.g. maximize offloading users, minimize system-wide overhead [10], minimize computing failure ratio). In this paper, we try to obtain both the minimizing overhead and maximizing offloading users, which can be formulated mathematically as follows:

$$\min_{O} \max_{I} (\sum_{i=1}^{N} O_{a_i}, \sum_{i=1}^{N} I_{a_i}) \tag{10}$$

S.t.

$C1 : O_{a_i} = \begin{cases} O_i^l, \ if \ a_i = 0 \\ O_i^o, \ if \ a_i > 0 \end{cases}$

$C2 : I_{a_i} = \begin{cases} 0, \ if \ a_i = 0 \\ 1, \ if \ a_i > 0 \end{cases}$

$C3 : O_i^l \geq O_i^o$

$C4 : \sum_{i=1}^{N} a_i \leq M$

$C5 : \Upsilon_i \geq t_i = \begin{cases} t_i^l, \ if \ O_{a_i} = O_i^l \\ (t_i^{o,t} + t_i^{o,e}, \ if \ O_{a_i} = O_i^o) \end{cases}$

$C6 : \forall i \in N, \ \forall a_i \in A$

The objective of the multi-object programming function is to minimize the system-wide overhead value and maximize the offloading users at the same time. However, solving such multi-dimensional scheduling problem here is an NP-hard [16].

In this case, we proposed a stochastic fair allocation algorithm, based on the minimum required resource blocks for each admitted user, to specifically illustrate the benefits of vehicular edge cloud computing framework in vehicular computation. Hence, we introduce some basic principles needed in our proposed algorithm.

**Definition 1:** For the purpose of illustrating the performance of computing offloading approach, the vehicular user can always benefit from computation offloading by having a lower individual computation overhead (i.e. $O_i^o \leq O_i^l$). Without losing the generality, the total offloading computation time consumption must less than the task time latency limit (i.e. $t_i^o \leq \Upsilon_i$).

**Definition 2:** When the individual overhead of local and offloading have the same value (i.e. $O_i^o = O_i^l$), we call it the equilibrium point. At the equilibrium point, the number of allocated resource blocks can be noted as equilibrium point amount $a_i^*$.

We can obtain the equilibrium point $a_i^*$ with equation (1)-(9):

$$O_i^l = O_i^o \implies \lambda_i^t t_i^l + \lambda_i^e e_i^l \implies a_i^* =$$

$$\frac{(\lambda_i^e \delta_i P_i + \lambda_i^t \delta_i)}{180\sigma \left(\lambda_i^t \left(\frac{\beta_i}{f_i^l} - \frac{\beta_i}{f_i^o}\right) - \lambda_i^e \tau_i + \lambda_i^e \varphi \beta_i\right) log_2\left(1 + \frac{P_i |H_{i,s}|^2}{N_0}\right)} \quad (11)$$

Because the resource block is the minimum unit for resource allocation scheme, we choose to round it up to the next integer of as the actual value in resource allocation scenario (i.e. $\lceil a_i^* \rceil$, where $\lceil \bullet \rceil$ denotes rounding up to the next integer function).

**Definition 3:** The allocated resource blocks amount must fulfill the requirement of time latency limit. The number of resource blocks can be noted as $a_i^\Delta$ when the time latency requirement can be just met.

We can get the $a_i^\Delta$ with equations (12) and (13):

$$\Upsilon_i = t_i^{o,t} + t_i^{o,e} \implies \Upsilon_i = \frac{\delta_i}{r^\Delta(a_i)} + \frac{\beta_i}{f_i^o} \quad (12)$$

$$a_i^\Delta = \frac{\delta_i f_i^o}{180\sigma(\Upsilon_i f_i^o - \beta_i) log_2(1 + \frac{P_i |H_{i,s}|^2}{N_0})} \quad (13)$$

Finally, we can obtain the minimum required resource block amount for user $i$ by equation (14).

$$a_i^{min} = \max_{i \in N} (\lceil a_i^* \rceil, \lceil a_i^\Delta \rceil) \quad (14)$$

As shown in equation (14), we can prove the existence of equilibrium point $a_i^*$ for each user. Since the resource blocks (RB) is the minimum unit for allocating. In this case, when the controller allocates at least $\lceil a_i^* \rceil$ RBs for user $i$, this user will begin to benefit from computing offloading. Simultaneously, the computation task i also need to be finished within its time latency limits, when the controller allocates at least $\lceil a_i^\Delta \rceil$ RBs for user $i$, this task can be executed successfully by offloading computation. Therefore, we can calculate the minimum amount of individual resource blocks by choosing the larger value between $\lceil a_i^* \rceil$ and $\lceil a_i^\Delta \rceil$.

The computation complexity of the SFA algorithm on eNodeB side is $O(n)$. Relatively, the computation complexity for each vehicular user is $O(n)$.

---

**Algorithm 1** Algorithm 1 Stochastic Fair Allocation (SFA) Algorithm _ eNodeB side

1: **Initialization:**
2: **Initialize** system remaining resource block $R(0)$.
3: **End initialization**

4: **while** $R(t) \geq \min_{i \in N}(a_i^{min})$ **do**
5:     **if** receive request for the available amount of resource block $R(t)$ **then**
6:         Send $R(t)$ to request users.
7:         **if** receive buffer state reports (BSRs) **then**
8:             **randomly choose** a request user $i$ and send a UL Grant to this user for offloading permission.
9:             **update** $R(t+1) = R(t) - a_i^{min}$ for the next slot.
10:         **else keep** $R(t+1) = R(t)$
11:         **end if**
12:     **else** continue listening requests
13:     **end if**
14: **end while**

---

**Algorithm 2** Algorithm 1 Stochastic Fair Allocation (SFA) Algorithm _ vehicular user side

1: **Initialization:**
2: **Initialize** each user individual decision profile.
3: **compute** the minimum required individual resource blocks based on equation (11)-(14).
4: **End initialization**

5: **while** $R(t) \geq a_i^{min}$ **do**
6:     **send** a buffer state report (BSR) to eNodeB requesting for $a_i^{min}$ resource blocks to offloading.
7:     **if** user $i$ receive the UL Grant of resource allocation **then**
8:         **update** $a_i(t+1) = a_i^{min}$ for the next slot.
9:     **else** keep $a_i(t+1) = 0$.
10:     **end if**
11: **end while**

## IV. NUMERICAL RESULTS

In this section, we conduct experiments to explore the serviceability and performance of VECC framework based SFA algorithm. The experiment parameters are listed in Table 1.

We firstly set the experimental cellular network bandwidth as 20 MHz, 30 vehicular users were randomly assigned tasks from 4 selected kinds of computation tasks. We record the response of edge cloud server during its decision process.

The effectiveness of Stochastic Fair Allocation Algorithm can be shown in Fig. 2 and Fig. 3. As we can see in Fig. 2, the changing process of the number of offloading users keeps in smooth linear growth until decision slot 28. At decision slot 28, the number of offloading users stops increasing, because the wireless network remaining resource block is not able to enable another user to offload its computing task. At last,

TABLE I: Experiment Parameters and Descriptions

| Symbols | Value | Units | Meaning | Supplement |
|---|---|---|---|---|
| $R$ | 50 | Meter | Base station service radius | |
| $N$ | 30 | Vehicle | Number of vehicular users | |
| $B_o$ | 10/15/20 | MHz | Wireless network bandwidth | 4G LTE regular service bandwidth |
| $R$ | 50/75/100 | Units | Number of resource blocks in different bandwidth conditions | |
| $\sigma$ | 5 | Carrier | 5 carriers aggregation | Our model involves the carrier aggregation technology |
| $q_i$ | 100 | mWatts | Transmission power of vehicular users | |
| $g_{i,s}$ | $l_{i,s}^{-\epsilon}$ | dB | Channel gain | $l_{i,s}$ denotes the distance between vehicular user i and base station s. We set the path loss factor $\epsilon = 2$ |
| $\theta$ | -100 | dBm | Background noise power | |
| $\delta_i$ | {1000, 2000, 5000, 10000} | kB | Size of computation input data | Randomly assign these 4 kinds of computation tasks to each vehicular user |
| $\beta_i$ | {100, 300, 1000, 2000} | Reference Supplement | Total required computation units | Megacycles in CPU computation or Gigaflop in GPU computation |
| $\Upsilon_i$ | {0.2, 0.6, 1, 2} | Second | Time delay limit | |
| $f_i^l$ | {0.5, 0.8, 1.0} | Reference Supplement | Vehicular local computation capacity | GHz in CPU computation or Teraflops in GPU computation |
| $\varphi$ | 0.0025 | Joule | Energy consumption per computation unit | |
| $f_i^o$ | 10 | Reference Supplement | Edge computation capacity allocated to user | GHz in CPU computation or Teraflops in GPU computation |
| $\lambda_i^t / \lambda_i^e$ | {0, 0.5, 1} | | Weight of time/energy consumption | $\lambda_i^t + \lambda_i^e = 1$ |

28 of 30 users successfully execute their tasks in VECC by offloading their task to edge cloud server, the other 2 left users execute their tasks by onboard computers. Fig. 3 shows the changing process of the number of VECC users with 4 types of computing tasks. These 4 kinds of tasks stay in a balanced opportunity to successfully execute their computing tasks on the edge cloud server, which reflects the fairness in this algorithm.

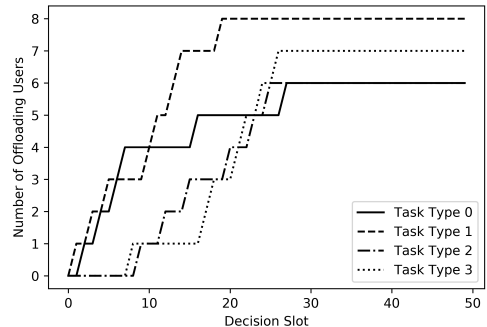

Fig. 3: The number of offloading users in 4 types of computing task in the decision process

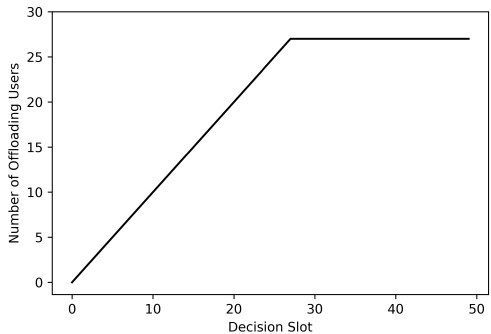

Fig. 2: The number of offloading users in the decision process

The VECC performance of the vehicular computing can be evaluated in Fig. 4 to Fig. 6. Fig. 4 shows the individual overhead changing process in VECC. Each colored line in this figure represents a single vehicular users overhead. It is easy to identify that most of the lines have a numerical reduction, which means the overhead of these vehicular users reduces in VECC. As further illustrated in Fig. 5. The system wide overhead, including both energy and time, reduces from 52.42 to 21.67. That means the overall computing overhead among these 30 vehicular users reduces around 58.66% in VECC. The energy only consumption reduces from 67.75 to 13.07, which decrease percentage reaches 80.71%. However, because we only allocate the minimum required amount of wireless resource blocks to each user, which will not reflect a great advantage on the reduction of time consumption. The time consumption reduces from 39.53 to 35.39, which reduces 10.47%. Moreover, in order to explore the performance and effectiveness of VECC and SFA algorithm, we conduct the experiment under 3 settings of general 4G bandwidth for 100 times, which includes 10 MHz, 15 MHz and 20 MHz. The results are illustrated in Fig. 6. The average number of offloading users in these 3 kinds of bandwidth cellular network are 21, 26, 29, separately. In 10 MHz bandwidth 4G cellular network, the system-wide overhead reduction, time only consumption reduction and energy only consumption reduction reach 52.18%, 11.52% and 79.03%, separately. In 15 MHz bandwidth 4G cellular network, these three evaluation parameters, arranged in order, reach 60.31%, 13.46% and 92.25%, separately. In 20MHz, they reach 61.48%, 14.61% and 93.76%, separately.

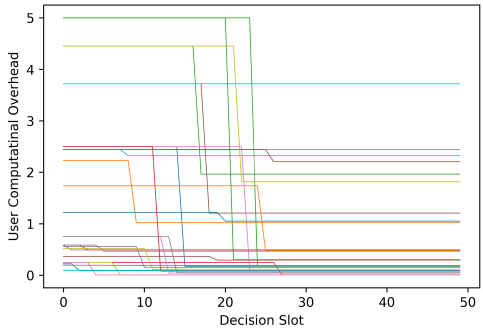

Fig. 4: The individual overhead in the decision process

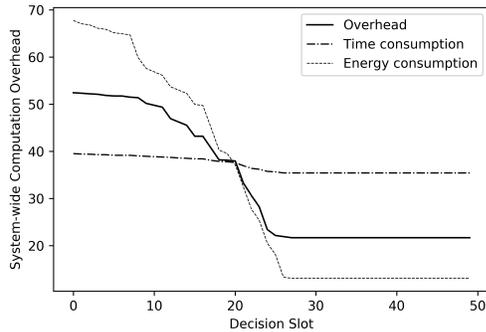

Fig. 5: The system wide computing overhead in the decision process (total overhead, time overhead and energy overhead)

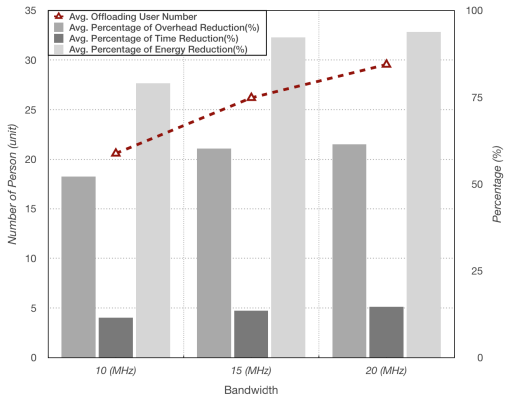

Fig. 6: The average performance of VECC in 10/15/20 MHz eNodeB bandwidth based SFA Algorithm

## V. CONCLUSION

In this paper, we proposed a vehicular edge cloud offloading (VECC) framework for offloading the intelligent vehicles local computations into the edge cloud. A stochastic fair allocation algorithm was proposed to vividly illustrate the benefits of the VECC. Eventually, the numerical results demonstrated the effectiveness of the VECC in reducing the onboard computational power consumption. In our future work, we will keep improving the framework and build applications under the VECC.

## ACKNOWLEDGMENT

The work is supported by the Doctoral Student Overseas Study Program No.201606370144 funded by the China Scholarship Council. The authors want to thank Professor Kun Zhou at the University of California, Berkeley for his insights on this research.